\documentclass[aps,twocolumn,nofootinbib,amsmath,amssymb,reprint,preprintnumbers,
]{revtex4-1}

\usepackage{dcolumn}
\usepackage[
	  pagebackref=false,
	  colorlinks=true,
      linkcolor=blue,
      urlcolor=blue,
      filecolor=black,
      citecolor=red,
      pdfstartview=FitV,
      pdftitle={},
        pdfauthor={},
        pdfsubject={},
        pdfkeywords={},
        pdfpagemode=None,
        bookmarksopen=true
      ]{hyperref}

\usepackage[normalem]{ulem}
\usepackage{amsmath}
\usepackage{enumerate}
\usepackage{amsfonts}
\usepackage{epsfig}
\usepackage{mathbbol}

\usepackage{color}
\usepackage{ulem}

\def\ep{\epsilon}
\newcommand{\be}{\begin{equation}}
\newcommand{\ee}{\end{equation}}
\newcommand{\bea}{\begin{eqnarray}}
\newcommand{\eea}{\end{eqnarray}}
\newcommand{\bega}{\begin{gather}}
\newcommand{\eega}{\end{gather}}
\newcommand{\nn}{\nonumber\\}
\newcommand{\bi}{\begin{itemize}}
\newcommand{\ei}{\end{itemize}}
\newcommand{\ben}{\begin{enumerate}}
\newcommand{\een}{\end{enumerate}}
\newcommand{\bca}{\begin{cases}}
\newcommand{\eca}{\end{cases}}
\newcommand{\bln}{\begin{align}}
\newcommand{\eln}{\end{align}}
\newcommand{\bst}{\begin{split}}
\newcommand{\est}{\end{split}}
\def\ie{\begin{equation}\begin{aligned}}
\def\fe{\end{aligned}\end{equation}}
\newcommand{\bma}{\le(\begin{matrix}}
\newcommand{\ema}{\end{matrix}\ri)}

\newcommand{\M}{\mathcal{M}}
\newcommand{\Mo}{\mathcal{M}_1}
\newcommand{\Mt}{\mathcal{M}_2}
\newcommand{\Mot}{\mathcal{M}_{1,2}}

\newcommand{\mg}[1]{\textcolor{black}{ #1}}

\begin{document}

\title{Novel shadows from the asymmetric thin-shell wormhole}
\author{Xiaobao Wang $^{1}$}
\email{bao@mail.bnu.edu.cn}

\author{Peng-Cheng Li$^{2,3}$}
\email{lipch2019@pku.edu.cn}

\author{Cheng-Yong Zhang$^4$}
\email{zhangcy@email.jnu.edu.cn}

\author{Minyong Guo $^{2}$}
\email{Corresponding author:minyongguo@pku.edu.cn}

\affiliation{$^1$ School of Applied Science, Beijing Information Science and Technology University, Beijing 100192, P. R. China\\
 $^2$Center for High Energy Physics, Peking University, No.5 Yiheyuan Rd, Beijing 100871, P. R. China\\
 $^3$Department of Physics and State Key Laboratory of Nuclear
Physics and Technology, Peking University, No.5 Yiheyuan Rd, Beijing
100871, P.R. China\\
$^4$ Department of Physics and Siyuan Laboratory, Jinan University,
Guangzhou 510632, China\\
}
\begin{abstract}

For dark compact objects such as black holes or wormholes, the shadow size has long been thought to be determined by the unstable photon sphere (region). However, by considering the asymmetric thin-shell wormhole (ATSW) model, we find that the impact parameter of the null geodesics is discontinuous through the wormhole in general and hence we identify novel shadows whose sizes are dependent of the photon sphere in the other side of the spacetime. The novel shadows appear in three cases: (A2) The observer's spacetime contains a photon sphere and the mass parameter is smaller than that of the opposite side; (B1, B2) there' s no photon sphere no matter which mass parameter is bigger. In particular, comparing with the black hole, the wormhole shadow size is always smaller and their difference is significant in most cases, which provides a potential way to observe wormholes directly through  Event Horizon Telescope with better detection capability in the future.


\end{abstract}
\maketitle

\noindent {\it Introduction.}--- The image of M87* taken by Event Horizon Telescope made its debut and revealed the shadow of black holes in April 2019 \cite{Akiyama:2019cqa}, which has ignited a wave of researches into the shadows of black holes. As is well-known, the shadow size of a black hole is considered to be determined by the unstable photon sphere (region) other than the event horizon from the earliest work \cite{Synge:1966okc,Bardeen:1973}.  As one of the most important predictions of general relativity (GR)  besides black holes, wormhole spacetime \cite{Visser:1995cc}  also contains unstable photon sphere which also creates shadows \mg{which were found to be different from the black holes in the size or the oblateness, see examples in   \cite{Sarbach:2012wi,Bambi:2013nla, Ohgami:2015nra, Nedkova:2013msa, Lamy:2018zvj, Vincent:2020dij}}. Even though it has been shown a traversable wormhole with ordinary matter is not allowed in GR, if there is some exotic matter in the universe, it becomes possible \cite{Ellis:1973yv,Morris:1988cz}. Furthermore, if the exotic matter distributes into a thin shell, a simple model of thin-shell wormhole can be constructed by the ``cut and paste'' technique \cite{Visser:1989kh, Visser:1989kg}. \mg{And as far as we know, the shadow of a asymmetric thin-shell wormhole has not been studied before in general.}

In this letter, we focus on an asymmetric thin-shell wormhole (ATSW) spacetime, more specially, a static thin shell connecting two distinct Schwarzschild spacetimes whose difference is only the mass parameter \cite{Visser:1989kg}. \mg{The Generic spherically symmetric dynamic thin-shell traversable wormholes and their stabilities were discussed in \cite{Garcia:2011aa}.} Using the backward ray-tracing method, we carefully analyze the ingoing null particles that emit from the observer' s position in one side and pass through the throat to the opposite side of the thin shell. Note that the static observers on both sides of the spacetime is different, thus we first find the impact parameter of the same photon defined in each side is different and related by a simple equation. Correspondingly, the turning point in each side for the same photon becomes different. Considering a physical process that  ingoing photons with no turning point would fall into the thin shell and then turn into outgoing in the other side, we first find some of them may hit its turning point defined in this spacetime and then turn back to its birthplace. Thus, we can see the difference between the traversable wormhole and black hole from the null geodesic motion. It's worth noting that for a black hole,  the light that enters the event horizon never returns, since the event horizon is a ``one-way'' membrane.

\begin{figure}[htbp]
\centering
\includegraphics[width=8.5cm]{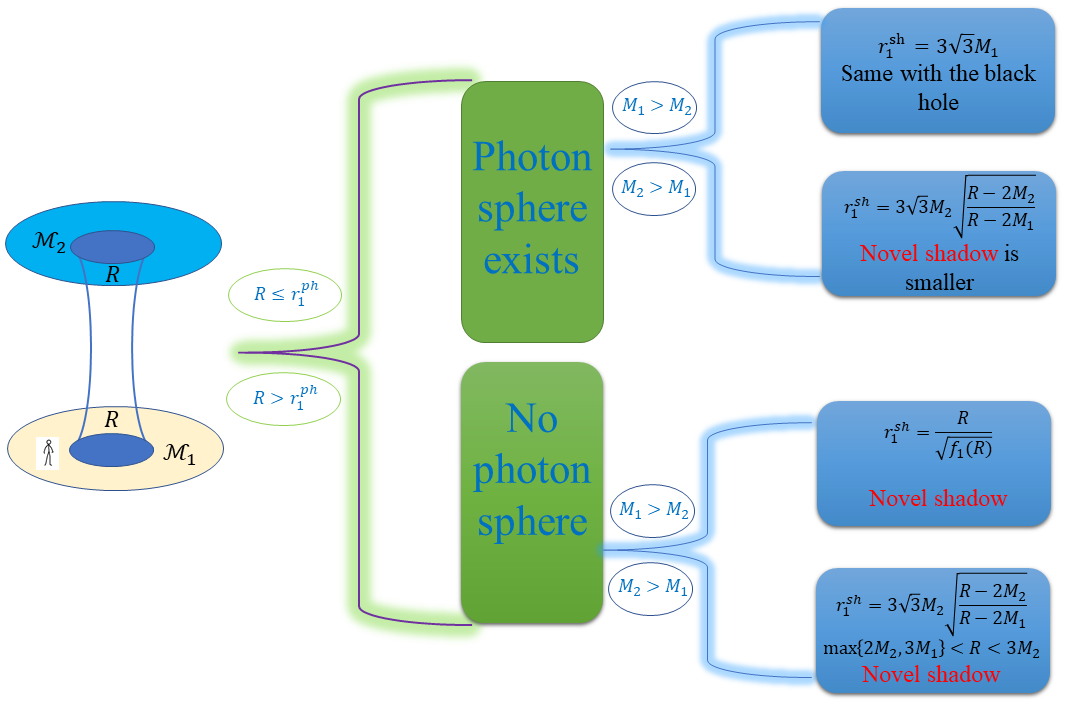}
\caption{The shadow of thin-shell wormhole for various parameters. There is novel shadow for some situations.}
\label{diag}
\end{figure}

Based on our new finding, we identify novel shadows in ATSW spacetime (see \cite{Dai:2019mse} for a similar study ). In particular, these novel shadows originate from spacetime asymmetry of both sides connected by the throat and their sizes do not depend on the photon sphere, contrary to what we typically think about. Specifically, let's assume we stay in the spacetime $\Mo$ and the radius of the thin shell is $R$. See Fig. \ref{diag}, the shadow of the wormhole seen by the static observer is the same with that of the corresponding black hole when $\Mo$ contains the photon sphere $R\le r_1^{ph}$ and $M_1>M_2$, that is, there is no novel shadow and the throat behaves like an ``event horizon'' . However, we find novel shadows in other cases. When $R\le r_1^{ph}$ and $M_2>M_1$, the shadow size of the wormhole becomes $r_1^{sh}=3\sqrt{3}M_2\sqrt{\frac{R-2M_2}{R-2M_1}}$ which is determined by the photon sphere of the opposite side and always smaller than that in Schwarzschild spacetime since the photons with the impact parameter $b_1<3\sqrt{3}M_1$ would fall into the spacetime $\Mt$ and turn back to $\Mo$ which never occurs for black hole. In particular, the size of the shadow can range from zero to $3\sqrt{3}M_1$, so that for some parameters the difference can be large enough to be detected to distinguish the wormhole from the black hole. When there is no photon sphere in $\Mo$, that is, $R>r_1^{ph}$, surprisingly, there also exists a novel shadow. For $M_1>M_2$ since $R$ is the turning point for certain null geodesics with the corresponding impact parameter $b_1^R$ and the photons with $b_1<b_1^R$ will go through the thin shell and never turn back, the shadow of the wormhole depends on the thin shell and its radius is $r_1^{sh}=\frac{R}{\sqrt{f_1(R)}}$. While for $M_2>M_1$ and $\max\{2M_2, 3M_1\}<R<3M_2$, the side $\Mt$ contains the photon sphere and  some null geodesics with $b_1<b_1^R$ could turn back after arriving at the spacetime $\Mt$, thus the critical impact parameter is $3\sqrt{3}M_2\sqrt{\frac{R-2M_2}{R-2M_1}}$ and the novel shadow size is $r^{sh}_1=3\sqrt{3}M_2\sqrt{\frac{R-2M_2}{R-2M_1}}$ which is smaller than $b_1^R$. Interestingly, this expression is the same with that for $M_2>M_1$ and $R\le r_1^{ph}$, see Fig. \ref{diag}.


\medskip

\begin{figure}[htbp]
\centering
\includegraphics[width=8.5cm]{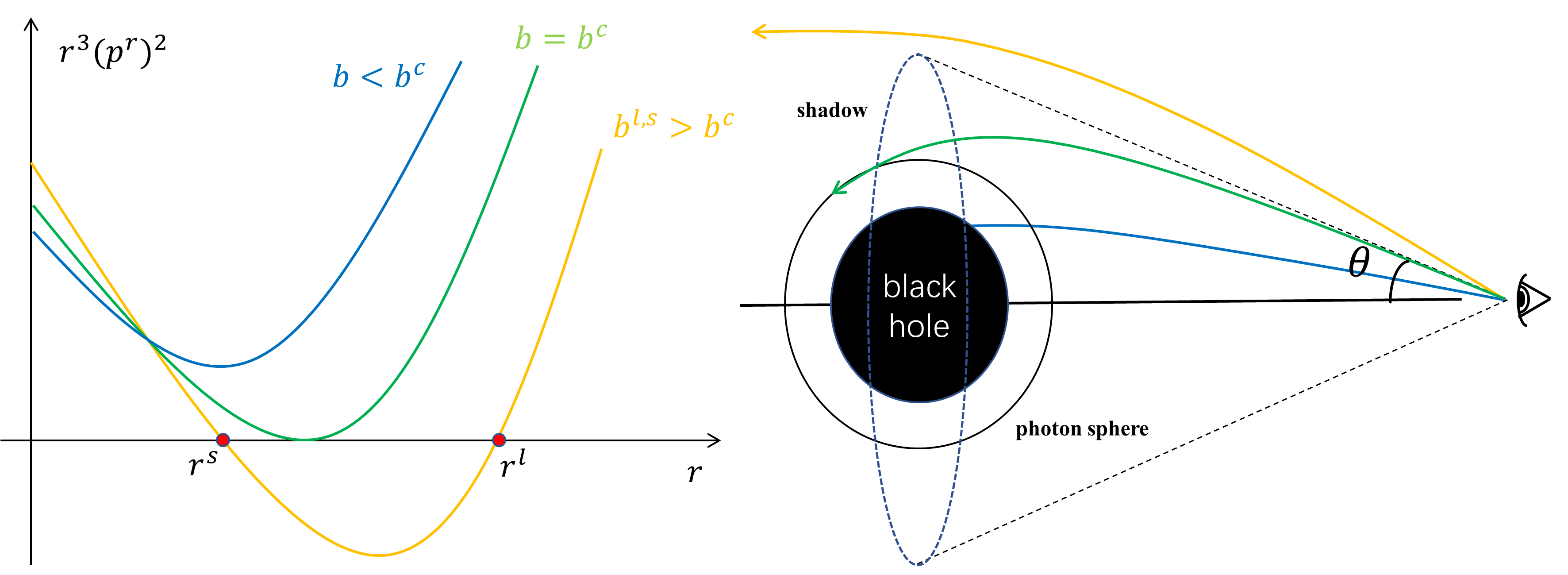}
\caption{The left picture shows behaviors of the function $r^3(p^r)^2$ with three different kinds of impact parameters. There's no turning point for $b<b^c$, single turning point for $b=b^c$ and two turning points for $b>b^c$. The right one presents the shadow of  Schwarzschild black hole.}
\label{blackhole}
\end{figure}

 \noindent {\it Background of Schwarzschild black hole shadow and the thin shell wormhole.}---Let us start from some important physical quantities and useful symbols in terms of the shadow in the background of Schwarzschild black hole whose metric takes in this form
\be
ds^2=-f(r)dr^2+f^{-1}(r)dr^2+r^2d\Omega^2,
\ee
where $f(r)=1-2M/r$. Considering a null particle with momentum vector $p^a$, the radial motion of the geodesic is described by the equation
\be\label{radialequation}
p^r=\pm E\sqrt{1-\frac{b^2}{r^2}f(r)},
\ee
where $b=L/E$ is the impact parameter and $\pm$ stands for outgoing and ingoing directions, respectively. The equation $p^r=0$ gives $b^2=r^2/f(r)$ has two real roots for the positive branch of $r$ when $b\ge b^c=3\sqrt{3}M$. For $b=b^c$, both roots are equal to $r^{ph}=3M$ is known as the radius of photon sphere. For $b>b^c$, the roots are addressed as the turning points with the larger root $r^l>r^{ph}$ and the smaller one $2M=r_h<r^s<r^{ph}$, where $r_h$ is the radius of the event horizon. Also, we introduce a new parameter $b^{l,s}\equiv r^{l,s}/\sqrt{f(r^{l,s})}$ that will be later used. Moreover, considering a distant static observer at $(t_o, r_o\to\infty, \pi/2, 0)$, the angular radius of the shadow left by the photon sphere satisfies $\sin\theta=b^c/r_o$ and the radius of the shadow is defined by $r^{sh}\equiv r_o \sin\theta=b^c$, as seen in Fig. \ref{blackhole}.

\begin{figure}[htbp]
\centering
\includegraphics[width=4.5cm]{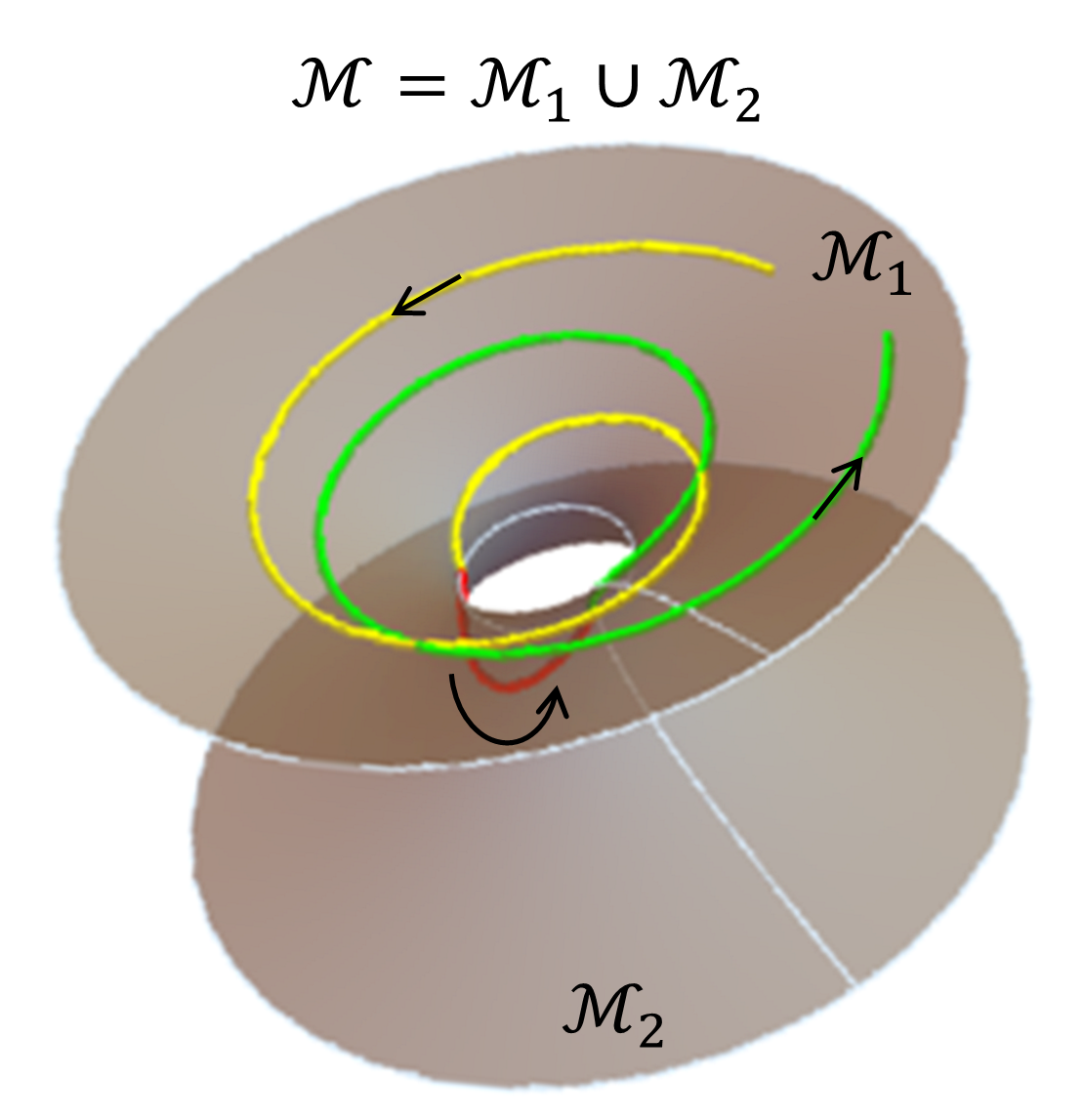}
\caption{The diagram of the thin-shell wormhole spacetime. In this figure, we give an example that in $\Mo$ an ingoing photon with certain impact parameter pass through the thin shell and stops at its turning point in $\Mt$ and turns back to $\Mo$.}
\label{wormhole}
\end{figure}

Next, let's introduce some necessary background knowledge of thin-shell wormhole using cut-and-paste method, that is, two distinct spacetimes $\Mot$ with different parameters are glued by a thin shell which forms a new manifold $\M=\Mo\cup\Mt$, as seen in Fig. \ref{wormhole}. We suppose that a spherical thin shell is moving in a spherically symmetric spacetime and the metrics on both sides take in this form
\be
ds_i^2=-f_i(r_i) dt_i^2+f_i^{-1}(r_i) dr_i^2+r_i^2d\Omega^2,
\ee
where $i=1,2$, and by focusing on the Schwarzschild case we have
$f_i(r)=1-\frac{2M_i}{r_i}$, where $M_{i}$ are the mass parameters.

The local tetrads in the neighbourhood of the thin shell of each spacetime $\Mot$
\begin{eqnarray}
\mathbf{e}_{t_i}^a&\equiv&f_i^{-\frac12}(R)\left(\dfrac{\partial}{\partial t_i}\right)^a\,,\\
\mathbf{e}_{r_i}^a&\equiv&\sqrt{f_i}(R)\left(\dfrac{\partial}{\partial r_i}\right)^a\,,
\end{eqnarray}
are related by the following
Lorentz transformation \cite{Langlois:2001uq}
\begin{equation}
\left(\begin{array}{c}
\mathbf{e}_{t_2}^a\\
\mathbf{e}_{r_2}^a
\end{array}\right)=\Lambda(\Theta_1-\Theta_2)\left(\begin{array}{c}
\mathbf{e}_{t_1}^a\\
\mathbf{e}_{r_1}^a
\end{array}\right)\,.
\end{equation}
where we have defined

\begin{equation}
\Lambda(\Theta)=
\left(\begin{array}{cc}
 \cosh{\Theta}$\quad $\sinh{\Theta}\\
\sinh{\Theta}$\quad $\cosh{\Theta}
\end{array}\right)\,,
\end{equation}
with
\be
\Theta_i=\sinh^{-1}\frac{\ep_i\dot{R}}{\sqrt{f_i(R)}}\,.
\ee
Here $R$ is the radius of the thin shell and $\dot{R}$ denotes the velocity of the moving thin shell.

\medskip

\noindent {\it Conserved quantities and some conventions}---Now, we consider an ingoing photon from the spacetime $\Mo$ passing through the thin shell. For simplicity, we assume the interaction between the photon and the thin shell is only governed by gravity which implies the 4-momentum $p^a$ is invariant during the process passing through the thin shell. \mg{The similar treatment has been applied in \cite{Nakao:2013hba}} In addition, the metric of the spacetime $\M$ is continuous by definition which means $g_{ab}^{\Mo}(R)=g_{ab}^{\Mt}(R)$. Hence, the conserved quantities $p_{t_i}=-E_i$ and $p_{\phi_i}=L_i$ along a geodesic imply
\bea\label{import}
-\frac{E_2}{\sqrt{f_2(R)}}&=&-\frac{\cosh\left(\Theta_1-\Theta_2\right)}{\sqrt{f_1(R)}}E_1+\frac{\sinh\left(\Theta_1-\Theta_2\right)}{\sqrt{f_1(R)}}p_R^{r_1},\nn
L_1&=&L_2,
\eea
at the position of the thin shell, and here we have used $p^{r_1}_R$ to denote the value of $p^{r_1}$ on the thin shell. Using Eq. (\ref{radialequation}), Eq. (\ref{import}) can be simplified as
\be\label{mostimportant}
\frac{b_1}{b_2}=\sqrt{\frac{f_2(R)}{f_1(R)}},
\ee
where for simplicity we have set $\dot{R}=0$ without losing the physics nature, that is, we assume the thin-shell wormhole is static which can be proven that if we tolerate the violation of null energy condition, we could always find the thin-shell wormhole to be static. From this equation, we find $b_1=b_2$ when $f_1(R)=f_2(R)$, that is $M_1=M_2$, which tells us that the spacetime $\M$ is symmetric about the thin shell. However, if $b_1\neq b_2$,  the mass parameters of $\Mo$ and $\Mt$ are no longer equal and the spacetime $\M$ becomes asymmetric about the thin shell which we focus on in this letter.

Before we move to discuss the shadow of the asymmetric thin-shell wormhole, in order to calculate conveniently we would like to establish some conventions. Let's suppose $M_1=1$ and $M_2=k>0$ for the mass parameter. Thus we have $R>\max\{2, 2k\}$ for thin-shell wormhole. Based on the previous review of the background of the shadow, the radius of the photon sphere of $\Mo$ and $\Mt$ is $r^{ph}_1=3$ and $r^{ph}_2=3k$, respectively. Correspondingly, the critical impact parameter reads $b_1^{c}=3\sqrt{3}$ and $b_2^{c}=3\sqrt{3}k$ for each side. Furthermore, as mentioned before we use $b^R_{1,2}=\frac{R}{\sqrt{f_{1,2}(R)}}$ to denote the specific impact parameter in which case one finds $p^R_{1,2}=0$. Using these conventions, Eq. (\ref{mostimportant}) can be rewritten as
\be
\frac{b_1}{b_2}=\sqrt{\frac{R-2k}{R-2}}=\sqrt{1-\frac{2(k-1)}{R-2}}.
\ee

\medskip

\noindent {\it The size of the thin-shell wormhole}---Now, we directly face the calculation of the thin-shell wormhole shadow. Unlike horizons of black holes, the throat of thin-shell wormhole is not fixed, instead, the thin-shell throat only needs to be larger than the horizon of the black hole in the corresponding spacetime. Thus, whether the photon sphere exists in $\Mo$ is unknown, that is, both $\max\{2k, 2\}<R\le3$ and $R>3$ are possible,  but if the latter is true, wormholes are even more different from black holes in terms of null geodesic's motion. Therefore, we would like to discuss the former first.

\textbf{(A1).} Since the relationship of size between $M_1$ and $M_2$ is not clear, but it matters a lot. We might as well consider $0<k<1<R/2\le3/2$ now. In addition, for the geodesics with $b_1\ge b_1^c$, the known results in Schwarzschild spacetime containing a black hole can be applied to the thin-shell wormhole. Next, we carefully discuss the ingoing photons with $b_1<b_{1}^c=3\sqrt{3}$, we find
\be
b_2<\frac{\sqrt{R-2}}{\sqrt{R-2k}}3\sqrt{3}\equiv B_2^c.
\ee
Furthermore, since $R>2>2k$, there are always null geodesics with $b_2^R$ such that $p_R^{r_2}=0$. From simple analysis, we find $b_2<B_2^c\le b_2^R$ is always true for $2<R\le 3$. Hence, when $R>3k$, $R$ is the larger real root making $p^R_{2}=0$  which implies all the outgoing null geodesics with $b_2<b_2^R$ starting from the thin shell will go to  the infinity in $\mathcal{M}_2$. In the range $0<k<1$, the region that doesn't satisfy the condition $R>3k$ is $2/3<k<1$ and $2<R<3k$, where $R$ is the smaller real root. In this situation, one expects the outgoing photons $b_2^c<b_2<b_2^R$ will stop at the turning point in the spacetime $\Mt$ and bounce back to the $\Mo$, thus the radius of a novel shadow of the thin-shell wormhole would be $r_1^{sh}=3\sqrt{3}k\sqrt{\frac{R-2k}{R-2}}\equiv B_1^c$. However, after some algebraic calculus, we find $B_2^c<b_2^c$ always holds for $2/3<k<1$ and $2<R<3k$, which tells us there is no turning point for null geodesics with $b_2<B_2^c$. From the above, we conclude that when $M_1>M_2$ and the side $\Mo$ contains a photon sphere, the shadow of the thin-shell wormhole observed by the static observer is the same with the corresponding black hole.

\begin{figure}[htbp]
\centering
\includegraphics[width=6.5cm]{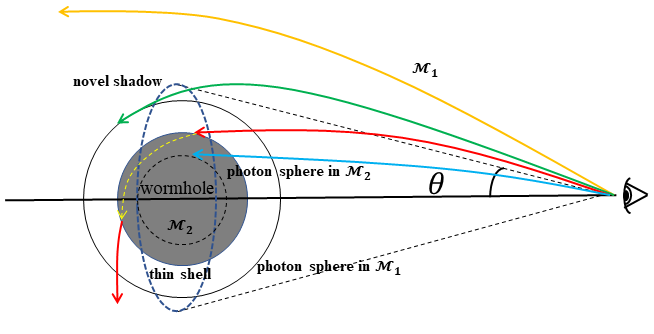}
\caption{The novel shadow for \textbf{(A2)} case, the spacetime $\Mo$ contains the photon sphere and the mass parameter of $\Mo$ is smaller than that of $\Mt$, namely, $1<k<R/2\le3/2$. It can be seen that the shadow of wormhole is smaller than that of black hole compared to the right graph in Fig. \ref{blackhole}. In fact, from our analysis, the novel shadow of wormhole is always smaller and could be arbitrarily small. }
\label{novelshadow}
\end{figure}

\textbf{(A2).}  Next, we turn to the case $b_1/b_2<1$, that is, $1<k<R/2$, we have $R\le3<3k=r_2^{ph}$ which implies the photon sphere always exists in the spacetime $\mathcal{M}_2$. Considering the ingoing null geodesics with $b_1<3\sqrt{3}$ in the spacetime $\mathcal{M}_1$, if we wish some geodesics would turn back passing through the throat, a necessary condition
\be
b_2=\sqrt{\frac{R-2}{R-2k}}b_1>b_2^c,
\ee
must be hold. Thus, we need
\be
B_1^c= \sqrt{\frac{R-2k}{R-2}}3\sqrt{3}k<b_1<b_1^c,
\ee
holds. Thus, the key is to check if this condition can be satisfied given $1<k<\frac{R}{2}$ and $2<R<3$. Note that $B_1^c$ is a decreasing function of $k$ in the range $k\in\left(1,\frac{R}{2}\right)$, we have $0<B_1^c<b_1^c$, that is to say, the desired condition is always satisfied. Therefore, we find that for $1<k<\frac{R}{2}\le3/2$, the angular radius of the shadow of the thin shell is always smaller than that of the corresponding black hole, see Fig. \ref{novelshadow}. In particular, for $R\sim 2k$, the radius of the shadow can be arbitrarily small. In terms of the novel shadow induced by the asymmetric mechanism of the thin-shell wormhole, the observational size of the shadow in the spacetime $\Mo$ can be very different between wormholes and black holes which may be detected by the EHT directly, if the resolution is raised to a good enough level.

\textbf{(B1).}  Now, we move to the case $R>3$, and also focus on $0<k<1$ at first. Since $R>3$, $R$ must be a outer turning point for a class of null geodesics with the impact parameter $b^R_1=\frac{R}{\sqrt{f_1(R)}}$. And the incident null geodesics will bounce back to the infinity when $b_1\ge b_1^R$ in the spacetime $\Mo$.

For $b_1<b_1^R$, note $b_1^R$ is an increasing function of $R$, we have $b_1^R>b_1^c$. When the null geodesics pass through the wormhole throat and arrive at the spacetime $\mathcal{M}_2$
 we have $b_2<\sqrt{\frac{R-2}{R-2k}}b_1^R=b_2^R$. Also, we find $b_2^R$ is an increasing function of $R$ for $0<k<1$, and obtain $b_2^R>b_2^c$. Combining with the condition $R>3k=r_2^{ph}$ we conclude the outgoing geodesics with $b_2<b_2^R$ will go to infinity in the spacetime $\mathcal{M}_2$, thus the novel shadow radius observed in the spacetime $\Mo$ is $r_1^{sh}=b_1^R$.

\textbf{(B2).}  Next, we pay our attention to the case $k>1$. Similarly, we find $b_2^R>b_2^c$ for $R>\max\{2k, 3\}$. Thus, for $R>3k$, the outgoing geodesics will go to infinity in the spacetime $\mathcal{M}_2$, that is, the shadow radius is $r_1^{sh}=b_1^R$. While, for $\max\{2k, 3\}<R<3k$, the critical impact parameter $b_2^c=3\sqrt{3}k$ corresponds to $B_1^c=3\sqrt{3}k\sqrt{\frac{R-2k}{R-2}}$ which is same with the case $2<R\le 3$ in surprise. And after some simple analysis, we find $B_1^c<b_1^R$ as expected. Therefore, in this case, the asymmetry of the spactime $\M$ results in a novel shadow very interestingly, which is also worth in-depth study.

\medskip

\noindent {\it Discussion.}---To summarize, by analyzing the behavior of null geodesics passing through the throat of the static thin-shell wormhole model, we identify novel shadows resulting from the asymmetry between the two sides of the thin shell and their sizes are not dependent on the photon sphere in the observer's spacetime. When the observer's spacetime contains a phono sphere, if a static observer lives on the side with a smaller mass parameter for a wormhole spacetime, taking the backward ray-tracing point of view, all the ingoing geodesics with impact parameters smaller than the critical one $3\sqrt{3}$ will pass through the thin shell and become outgoing in the other side of the spacetime. However, different from the ``one way'' property of black holes, some of them above a new critical impact parameter will stop at the turning point, bounce back to the original spacetime and come into the eyes of the static observer due to the asymmetry. Thus, the novel shadow is always smaller than the analogue of black holes. The difference can be very significant to overcome the  uncertainty in the measurements of black hole mass and its distance from observers so that  observing the thin-shell wormholes directly through Event horizon Telescope becomes feasible. While, the observer lives on the side with a larger mass parameter for a wormhole spacetime, the observational shadow is as same as that of a black hole,  so one cannot discern them by imaging shadow.

 Moreover, when the radius of the thin shell is larger than that of the photon sphere, we also show that there are novel shadows even though the observer's spacetime does not contain the photon sphere, which is an important complement to the related existing knowledge.

The mechanism for predicting this novel shadow of the thin-shell wormhole is very simple and elegant, we believe it could be a universal property. Thus it would be extremely interesting to see if this novel shadow can be found in other wormhole models in general relativity or modified theories of gravity. Furthermore, searching for this novel shadow by Event Horizon Telescope is also an exciting task in the future to know whether the thin-shell wormhole exists, or not. Concerning the EHT2017 observations \cite{Akiyama:2019cqa}, as analyzed in \cite{Akiyama:2019fyp} the compact objects whose shadows exhibit qualitatively deviation from those of black holes can be ruled out for M87* (however, see \cite{Bambi:2019tjh} for exception). So the thin-shell wormhole models can be strongly constrained with the present observations. However it may still has the chance to discover wormhole by EHT according to the novel shadow feature in the future.

\medskip

The work is in part supported by NSFC Grant No. 11335012, No. 11325522 and No. 11735001. MG and PCL are also supported by NSFC Grant No. 11947210. And MG is also funded by China Postdoctoral Science Foundation Grant No. 2019M660278 and 2020T130020. PCL is also funded by China Postdoctoral Science Foundation Grant No. 2020M670010. CYZ is supported by NSFC Grant No. 11947067.


\end{document}